\documentclass[final]{aipproc}

\layoutstyle{6x9}
\usepackage{amssymb}
\usepackage[colorlinks=TRUE, dvips]{hyperref}

\newcommand{\mnras}{MNRAS}
\newcommand{\aap}{A\&A}
\newcommand{\apj}{ApJ}

\newcommand{\apjl}{ApJL}
\newcommand{\aj}{AJ}

\def\msun{\hbox{${\rm M}_{\odot}$}}
\def\rsun{\hbox{${\rm R}_{\odot}$}}
\def\mstar{\hbox{$M_{\star}$}}
\def\rstar{\hbox{$R_{\star}$}}
\def\vsini{\hbox{$v\sin i$}}

\def\Prot{\hbox{$P_{\rm rot}$}}

\def\dOm{\hbox{$d\Omega$}}
\def\dOmsun{\hbox{$\dOm_\odot$}}
\def\chisq{\hbox{$\chi^2$}}
\def\kms{\hbox{km\,s$^{-1}$}}

\def\degr{\hbox{$^\circ$}}
\def\mrpd{\hbox{mrad\,d$^{-1}$}}

\begin{document}

\title{Large-scale magnetic topologies of M dwarfs}


\author{J.~Morin}{
  address={LATT, Universit\'e de Toulouse, CNRS, 14 Av.\ E.~Belin,
F--31400 Toulouse, France\\}
}

\author{J.-F.~Donati}{
  address={LATT, Universit\'e de Toulouse, CNRS, 14 Av.\ E.~Belin,
F--31400 Toulouse, France\\}
}

\author{X.~Delfosse}{
  address={LAOG--UMR~5571, CNRS et Univ.\ J.~Fourier, 31 rue de la
Piscine, F--38041 Grenoble, France\\}
}

\author{T.~Forveille}{
  address={LAOG--UMR~5571, CNRS et Univ.\ J.~Fourier, 31 rue de la
Piscine, F--38041 Grenoble, France\\}
}

\author{M.M.~Jardine}{
  address={SUPA, School of Physics and Astronomy, Univ. of St Andrews, St
Andrews, Scotland KY16 9SS\\}
}

\begin{abstract}
We present here the first results of a spectropolarimetric analysis of a
small sample ($\sim20$) of active stars ranging from spectral type M0 to
M8, which are either fully-convective or possess a very small radiative
core. This study aims at providing new constraints on dynamo processes in
fully-convective stars. 

Results for stars with spectral types M0-M4 -- i.e. with masses above or
just below the full convection threshold ($\simeq 0.35~\msun$) -- are
presented. Tomographic imaging techniques allow us to reconstruct the
surface magnetic topologies from the rotationally modulated time-series of
circularly polarised profiles.

We find strong differences between partly and fully convective stars
concerning magnetic field topology and characteristic scales, and
differential rotation. Our results suggest that magnetic field generation
in fully convective stars relies on different dynamo processes than those
acting in the Sun and other partly convective stars, in agreement with
theoretical expectations.

\end{abstract}

\maketitle

\section{Context}

In partly convective stars such as the Sun, magnetic fields -- the
energy source of most activity phenomena -- are induced by plasma motions:
the combined action of differential rotation ($\Omega$ effect) and
cyclonic convection ($\alpha$ effect) manages to generate a self-sustained
magnetic field. The so-called $\alpha\Omega$ dynamo \cite{Parker55} is
believed to operate mostly through the tachocline, a thin zone of strong
shear located at the interface between the radiative inner zone and the
convective envelope \cite{Charbonneau05}.

Stars with masses lower than about $0.35~\msun$ are fully-convective
\cite{Chabrier97} and thus do not possess a tachocline. However, they
manage to trigger magnetic fields \cite{Saar85,Johns96,Reiners06} and are
very active \cite{Delfosse98,Mohanty03,West04}. Though significant
progress was made since first non-solar dynamo mechanisms were proposed
\cite{Durney93}, theoretical and numerical modelling require observational
constraints. It is now acknowledged, from observational
\cite{Donati06,Morin08} and theoretical points of view
\cite{Chabrier06,Dobler06,Browning08}, that fully convective stars (FCS)
manage to yield large scale magnetic fields. But the properties of such
magnetic fields, and their dependency on stellar parameters (in particular
mass and rotation rate) are not yet clear.

We present here the first results of a spectropolarimetric analysis of a
small sample of active M dwarfs with spectral types ranging from M0 to M8,
which are either fully convective or possess a very small radiative core.
We aim at exploring the properties of the large-scale magnetic topologies 
of FCS, and their evolution with main stellar parameters. The stars were
selected from the rotation-activity study \cite{Delfosse98}. We chose only
active stars so that the magnetic field is strong enough to produce
detectable circularly polarised signatures, allowing us to apply
tomographic imaging techniques. More details about the sample are
available in Table~\ref{tab:sample}.

\begin{table}
\begin{tabular}{cccccccccccc}
 \hline
 Name & ST & \mstar & log$R_X$ & \vsini & \Prot & $\tau_c$ & $Ro$ &
\rstar & $i$ \\
 & & (\msun) & & (\kms) & (d) & (d) & ($10^{-2}$) & (\rsun) & (\degr) \\ 
 \hline
GJ 182 & M0.5 & 0.75 & $-3.1$ & 10 & 4.35 & 25 & 17 & 0.82 & 60 \\
DT Vir & M0.5 & 0.59 & $-3.4$ & 11 & 2.85 & 31 & 9.2 & 0.53 
& 60 \\
DS Leo & M0 & 0.58 & $-4.0$ & 2 & 14.0 & 32 & 44 & 0.52 & 60
\\
GJ 49 & M1.5 & 0.57 & $< -4.3$ & 1 & 18.6 & 33 & 56 & 0.51 & 45 \\
OT Ser & M1.5 & 0.55 & $-3.4$ & 6 & 3.40 & 35 & 9.7 & 0.49 &
45
\\
CE Boo & M2.5 & 0.48 & $-3.7$ & 1 & 14.7 & 42 & 35 &
0.43 & 45 \\
 AD Leo & M3 & $0.42$ & -3.18 & $3.0$ & $2.24$ & 48 & 4.7 & 0.38 &
20\\
 EQ Peg A & M3.5 & $0.39$ & -3.02 & $17.5$ & $1.06$ & 54 & 2.0 &
0.35 & 60 \\
 EV Lac & M3.5 & $0.32$ & -3.33 & $4.0$ & $4.37$ & 64 & 6.8 & 0.30
& 60 \\
 YZ CMi & M4.5 & $0.31$ & -3.09 & $5.0$ & $2.78$ & 66 & 4.2 & 0.29
& 60 \\
 V374~Peg & M4 & $0.28$ & -3.20 & $36.5$ & $0.446$ & 72 & 0.62 &
0.28 & 70 \\
 EQ Peg B & M4.5 & $0.25$ & -3.25 & $28.5$ & $0.404$ & 76 & 0.53 &
0.25 & 60 \\
 \hline
\end{tabular}
\caption{Fundamental parameters of the stellar sample. Columns 1--4
respectively list the name, the spectral type, the stellar mass, and the
logarithmic relative X-ray luminosity log$R_X=$log($L_X/L_{bol}$). The
projected rotation velocity and rotation period inferred from Zeeman
Doppler Imaging (ZDI) are mentioned in columns 5 and 6. Columns 7--10
respectively list the empirical convective turnover time from
\cite{Kiraga07}, the effective Rossby number $Ro=\frac{\Prot}{\tau_c}$,
the theoretical radius suited to the stellar mass from \cite{Baraffe98},
and the inclination angle used for ZDI. See \cite{Morin08b,Donati08} for
more details.}
\label{tab:sample}
\end{table}

For this study, we used the twin instruments ESPaDOnS on the 3.6-m
Canada-France-Hawaii Telescope (CFHT) located in Hawaii and NARVAL on the
2-m T\'elescope Bernard Lyot (TBL) in southern France. These
spectropolarimeters, built on the same design, can produce Stokes I,Q,U
and V spectra spanning the entire optical domain (from 370 to 1000~nm) at
a resolving power of $\sim 65000$ \cite{Donati03}. 

We performed monitoring observations of the sample in circularly polarised
light (Stokes $V$). Least-squares deconvolution (LSD) \cite{Donati97} was
then applied, resulting, for each spectra, in a synthetic line profile
gathering polarimetric information from most photospheric atomic lines.

\section{Imaging procedure and Model description}
For each star of the sample, our aim is to infer the topology of the
surface magnetic field from the circularly polarised (Stokes $V$) LSD
profiles we obtained. This can be achieved using a Zeeman-Doppler
Imaging (ZDI) code \cite{Donati97b}. The imaging process is based on the
principles of maximum entropy image reconstruction.  The magnetic field
is  decomposed into its poloidal and toroidal components, both expressed
as spherical harmonics expansions. Starting from a null magnetic field, we
iteratively improve our magnetic model by comparing the synthetic Stokes
$V$ profiles with the observed LSD profiles, until we reach an optimal
field topology that reproduces the data at a given \chisq\ level. The
inversion problem being partly ill-posed, we use the entropy function to
select the magnetic field with lowest information content among all those
reproducing the data equally well \cite{Skilling84}.

To compute the synthetic Stokes $V$ profiles, the star is
divided into a grid of $\sim 1000$ cells on which the magnetic field
components are computed directly from the coefficients of the spherical
harmonics expansion. The contribution of each individual pixel is
computed from a model based on Unno-Rachkovsky's equations
\cite{Morin08b}. We then integrate all contributions from the visible
hemisphere at each observed rotation phase.

While computing the Stokes $V$ profiles it is possible to account for
differential rotation. For a given differential rotation law, each local
line profile is Doppler-shifted as a function of the observation phase. It
is then possible to investigate how the fit quality varies in a reasonable
range of \Prot\ and \dOm\ values. We can thus derive the optimal \Prot,
\dOm\ and corresponding error bars \cite{Petit02,Donati03b,Morin08}.

\section{Reconstructed magnetic fields}
For each star of our sample, we reconstruct with ZDI the large-scale
surface magnetic field at least up to degree $\ell=6$. We also measure
differential rotation and assess time-variability of the magnetic topology
whenever possible.

\begin{figure}
\label{fig:plotMP}
  \includegraphics[height=.6\textwidth]{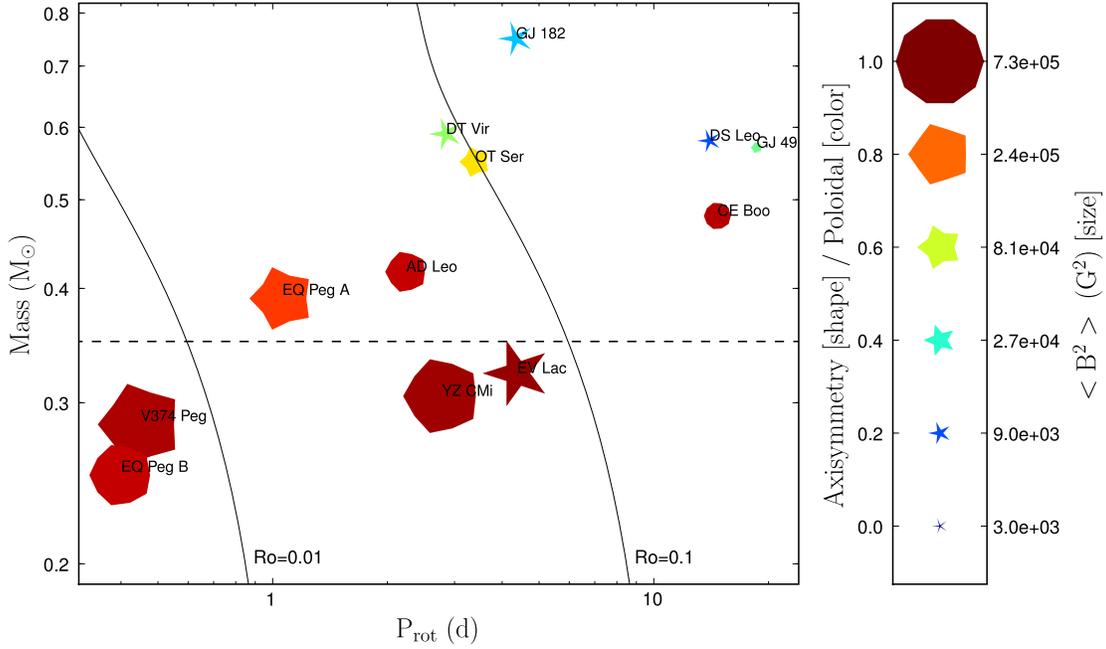}
  \caption{Properties of the magnetic topologies of M dwarfs as a
function of rotation period and stellar mass. Larger symbols indicate
larger magnetic fields while symbol shapes depict the different degrees
of axisymmetry of the reconstructed magnetic field (from decagons for
purely axisymmetric fields to sharp stars for purely non axisymmetric
fields). Colours illustrate the field configuration (dark blue for purely
toroidal fields, dark red for purely poloidal fields and intermediate
colours for intermediate configurations). Solid lines represent contours
of constant Rossby number $Ro=0.1$ and $0.01$. The theoretical
full-convection limit ($\mstar \simeq0.35\msun$ \cite{Chabrier97}) is
plotted as a horizontal dashed line.}
\end{figure}

The field is characterised by three quantities: (a) the overall magnetic
energy, (b) the ratio of magnetic energy reconstructed in the poloidal
modes and (c) the ratio of poloidal magnetic energy reconstructed in the
axisymmetric modes (defined by $m < \frac{\ell}{2}$). The results are
described below and presented in a more visual way in
Figure~\ref{fig:plotMP} as a function of \mstar\ and \Prot . See
[Donati, these proceedings] for a version of this diagram including G and
K dwarfs. To compare magnetic field generation in stars of different
masses, it is convenient to introduce the effective Rossby number which
rescales \Prot\ by a mass dependant coefficient.
$Ro=\frac{\Prot}{\tau_c}$, where $\tau_c$ is an empirical convective
turnover time inferred from X-ray luminosities \cite{Kiraga07}. On
Fig.~\ref{fig:plotMP} we also plot contours
of constant Rossby number $Ro=0.1$ and $0.01$ respectively
corresponding approximately to the saturation and super-saturation
thresholds \citep[e.g.,][]{Pizzolato03}.

\begin{figure}
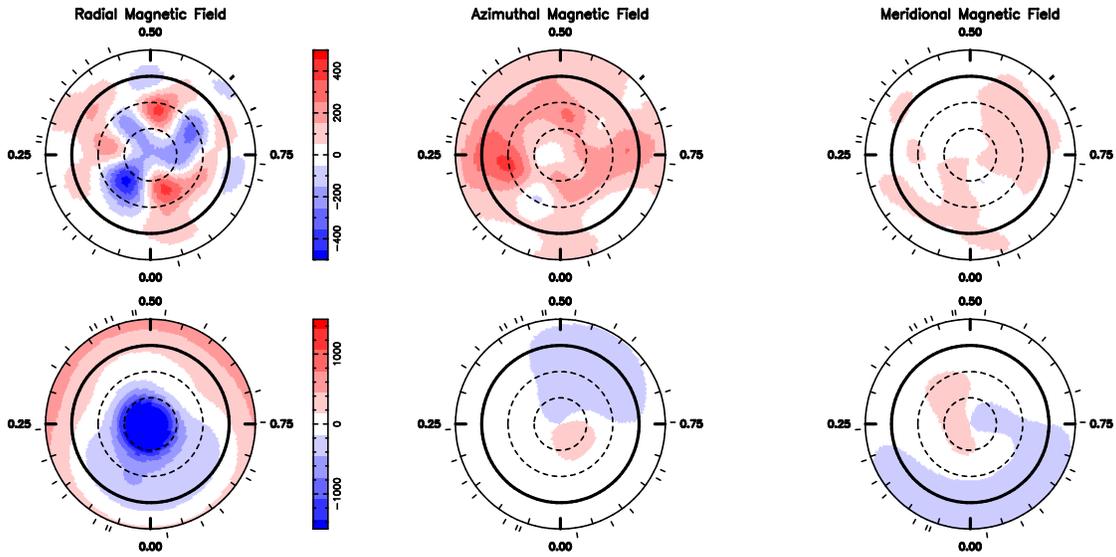

\begin{tabular}{c}
 \includegraphics[width=1.\textwidth]{mdw_494map2} \\
 \includegraphics[width=1.\textwidth]{yzcmi_V_08c_map} \\
\end{tabular}
\caption{\textbf{Upper row}: surface magnetic flux of DT~Vir (0.59~\msun)
as derived from our 2008 data set. The three components of the field in
spherical coordinates are displayed from left to right (flux values
labelled in G). The star is shown in flattened polar projection down to
latitudes of $-30\degr$, with the equator depicted as a bold circle and
parallels as dashed circles. Radial ticks around each plot indicate phases
of observations. \textbf{Bottom row}: same for YZ~CMi (0.31~\msun).}
\label{fig:zdi_maps}
\end{figure}

The magnetic field topologies we reconstruct with ZDI and the differential
rotation  amplitudes are very different on each part of the
$\mstar\simeq0.5\msun$ boundary. This threshold is very sharp and well
defined, with little apparent dependence with the rotation period. Typical
examples of magnetic maps on each part of this limit are shown in
Fig.~\ref{fig:zdi_maps}.

For stars more massive than $0.5~\msun$, we recover magnetic topologies
including (i) a \textbf{strong toroidal component} and (ii) a \textbf{high
non-axisymmetric degree of the poloidal component}. For 4 stars of this
subsample we can derive differential rotation. We find $\mathbf{\dOm
\gtrsim \dOmsun}$, in agreement with the dispersion observed in previous
photometric measurements of \Prot. As a consequence, \textbf{surface
magnetic features are short-lived}, the topology completely changes on a
timescale of a few months.

In the low mass subsample (\mstar<$0.5~\msun$), we reconstruct
\textbf{much stronger magnetic fluxes} and very different large-scale
magnetic topologies : (i) \textbf{mostly poloidal} ($\sim 90\%$ of the
reconstructed energy), (ii) \textbf{strongly axisymmetric} (except for
EV~Lac, more than half of the magnetic energy is reconstructed in $m=0$
modes) and (iii) \textbf{close to a dipole} with more 50\% of the
reconstructed magnetic energy lying in poloidal modes of degree $\ell=1$. 
\textbf{Very weak differential rotation} is inferred from our data with 3
stars having \dOm\ of the order of a few \mrpd. This very weak
differential rotation is in agreement with the most recent numerical 
simulations \cite{Browning08}. These stars were observed at two different
epochs separated by $\sim 1~{\rm yr}$. \textbf{Evolution of the magnetic
topologies is small}, in some cases, it is possible to fit
observations
separated by $\sim 1~{\rm yr}$ with a unique magnetic topology.

\section{Conclusions and perspectives}
\begin{figure}
 \label{fig:Ro_Bsq_RX}
 \begin{tabular}{cc}
  \includegraphics[width=.5\textwidth]{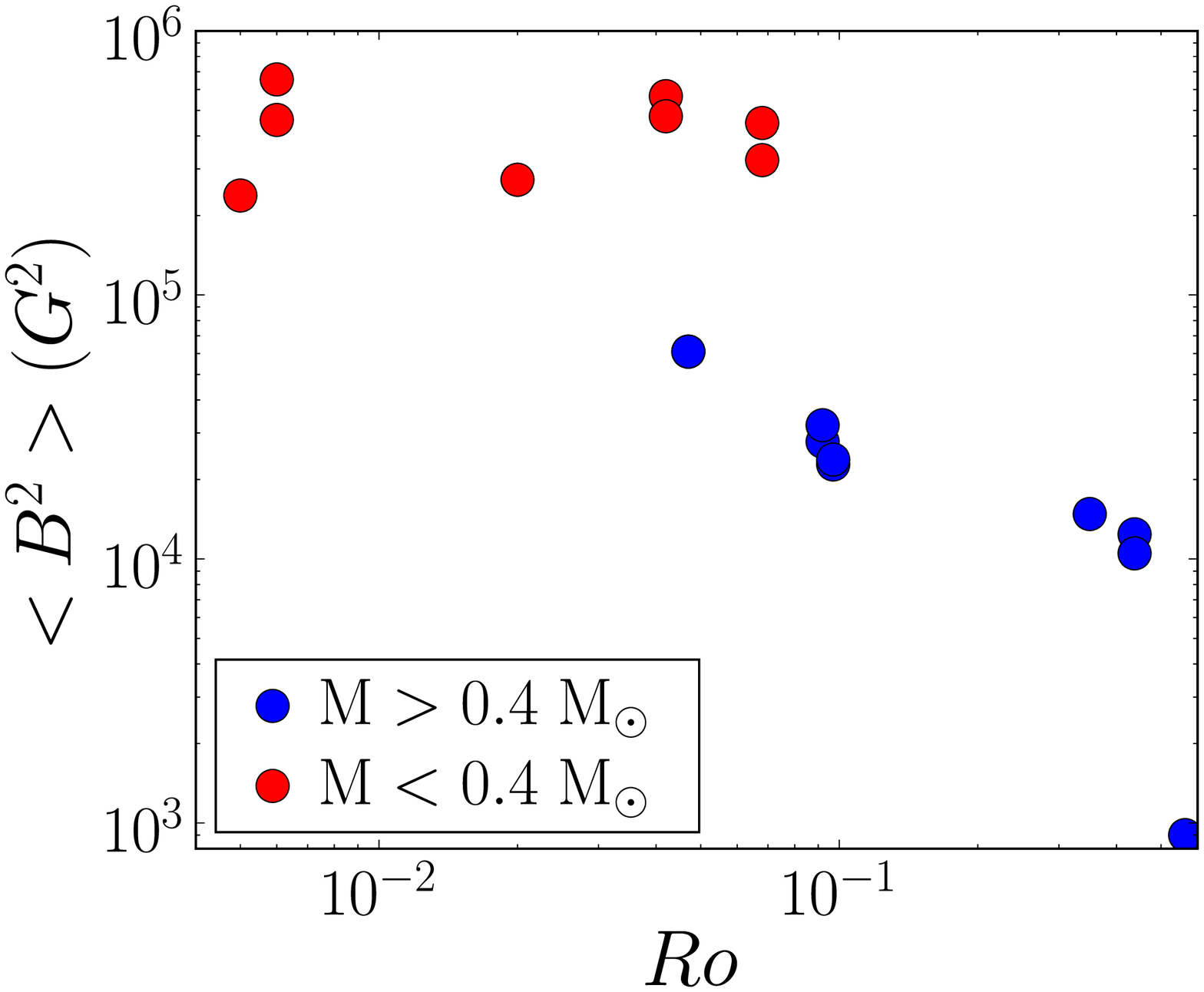} &
  \includegraphics[width=.5\textwidth]{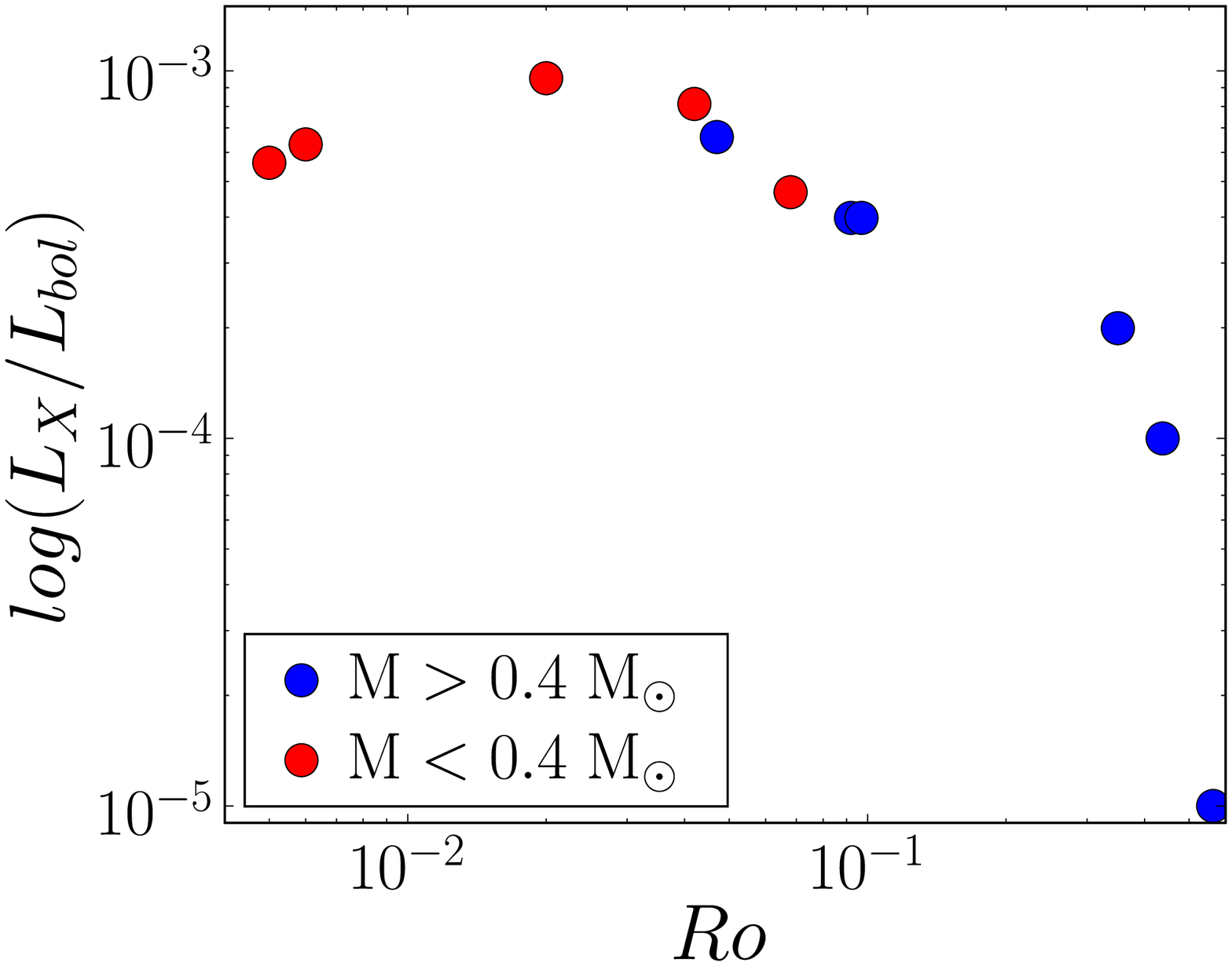} \\
 \end{tabular}
 \caption{Left panel: Reconstructed magnetic energy as a function of $Ro$.
Right panel: logarithmic relative X-ray luminosity as a function of $Ro$
(see Tab.~\ref{tab:sample}).}
 \end{figure}

Very different large-scale magnetic topologies are observed on each part
of the $\mstar\simeq0.5~\msun$ limit. We also note that dynamo processes
become suddenly much more efficient at triggering large-scale magnetic
fields (see Fig.~\ref{fig:plotMP} and \ref{fig:Ro_Bsq_RX}) at
approximately the same mass ($\simeq0.4~\msun$). This strong step is not
visible in the log$R_X$ vs $Ro$ plot. This result suggests that (i) the
X-ray emission is sensitive to overall magnetic energy whereas we are only
sensitive to the largest scales. (ii) At a given $Ro$ stars with mass
above or below $\sim 0.5~\msun$ generate comparable magnetic energy, but
with very different spatial scales repartition (the less massive stars
triggering more magnetic energy in the largest scales).

These strong changes in magnetic field generation occur at masses
slightly larger than the theoretical limit to full convection ($\simeq
0.35~\msun$). This may be due to strong shrinking of the radiative inner
zone predicted by theoretical models, from $\sim 0.5~\rsun$ at
$\mstar=0.5~\msun$ to nothing at $0.35~\msun$ \cite{Baraffe98,Siess00}.

We are currently completing this survey to extend it to fast rotators
with $\mstar>0.5~\msun$, slow rotators with $0.2<\mstar<0.5~\msun$, two
regimes not explored in the present sample. Present efforts are
also directed to ultracool dwarfs ($\mstar<0.2~\msun$), to investigate
how dynamo processes operate down to the brown dwarf limit, i.e. when
stellar atmospheres start to become neutral.

\begin{theacknowledgments}
  Julien Morin thanks the SOC of Cool Stars 15 and CNRS for providing
financial support for attending the conference.
\end{theacknowledgments}

\bibliographystyle{aipprocl} 

\hyphenation{Post-Script Sprin-ger}

\end{document}